\let\new=\newcommand
\new{\eq}{\begin{equation}}
\new{\en}{\end{equation}}
\new{\degree}{$^\circ$ }
 
\documentclass[iop]{emulateapj} 

\shorttitle{Hierachical multiples}
\shortauthors{Roberts et al.}

\begin{document}

%\received{ } 
%\accepted{ } 

\title{Observations of Hierarchical Solar-Type Multiple Star Systems}

\author{
Lewis C. Roberts, Jr.\altaffilmark{1}, 
Andrei Tokovinin\altaffilmark{2},
Brian D. Mason\altaffilmark{3}, 
William I. Hartkopf\altaffilmark{3},
Reed L. Riddle\altaffilmark{4} 
} 

\altaffiltext{1}{Jet Propulsion Laboratory, California Institute of Technology, 4800 Oak Grove Drive, Pasadena CA 91109, USA}
\altaffiltext{2}{Cerro Tololo Inter-American Observatory, Casilla 603, La Serena, Chile}
\altaffiltext{3}{U.S. Naval Observatory, 3450 Massachusetts Avenue, NW, Washington, DC 20392-5420, USA}
\altaffiltext{4}{Division of Physics, Mathematics, and Astronomy, California Institute of Technology, Pasadena, CA 91125, USA}
\email{lewis.c.roberts@jpl.nasa.gov}
 
%%%%%%%%%%%%%%%%%%%%%%%%%%%%%%%%%%%%%%%%%%%%%%%%%%%%%%%%%%%%%%%

\begin{abstract}
Twenty multiple stellar systems with solar-type primaries were observed at high angular resolution using the PALM-3000 adaptive optics system at the 5 m Hale telescope. The goal was to complement the knowledge of hierarchical multiplicity in the solar neighborhood by confirming recent discoveries by the visible Robo-AO system with new near-infrared observations with PALM-3000. The physical status of most, but not all, of the new pairs is confirmed by photometry in the $Ks$ band and new positional measurements. In addition, we resolved for the first time five close sub-systems: the known astrometric binary in HIP~17129AB, companions to the primaries of HIP~33555, and HIP~118213, and the companions to the secondaries in HIP~25300 and HIP~101430. We place the components on a color-magnitude diagram and discuss each multiple system individually.
\end{abstract}

 
%%%%%%%%%%%%%%%%%%%%%%%%%%%%%%%%%%%%%%%%%%%%%%%%%%%%%%%%%%%%%%%

\section{INTRODUCTION}
 
Statistics of binaries and hierarchical stellar systems trace conditions of star formation and serve as an excellent diagnostic for testing theoretical predictions and numerical simulations. Only recently have observational techniques such as adaptive optics (AO) and radial velocities (RV) reached the maturity and productivity needed for a high quality binary census over a wide range of periods. The latest results on binary statistics are reviewed by \citet{DK13}.

Reaching completeness for triple and higher-order hierarchies is even more challenging. The fraction of known  hierarchical systems in the nearby 25-pc volume has doubled recently \citep{R10} in comparison with the earlier work by \citet{DM91}. Large space volume and large samples are needed for their statistical study as the fraction of triples is only 14\% \citep{tokovinin2014b}. However, observing thousands of targets by complementary techniques would require prohibitive amounts of telescope and calendar time unless an intelligent strategy is applied. As noted by \citet{R10}, new observations mostly convert binaries into triples. So, focusing on known binaries and quantifying the presence of additional components (hence hierarchies) is a productive approach. It was adopted in the massive multiplicity survey of solar-type stars conducted in 2012--2013 with the Robo-AO instrument at the Palomar 60-inch telescope \citep{riddle2015}; hereafter R15. That survey looked for subsystems in the faint secondary components and for distant tertiary companions to known  binaries with FG primaries within 67 pc. The full sample is described in \citep{tokovinin2014a}.

We report here the beginings of follow-up observations to the R15 survey at the Palomar 5 m telescope. Our goal is to both confirm the newly discovered subsystems with infrared (IR) imagery and to combine that IR imagery with the Robo-AO observations in $i'$ band. Multi-color photometry allows us to place the secondaries on a color-magnitude diagram (CMD) to test their physical relation to the main target. For systems with high proper motion (PM), the second-epoch astrometry within a year allows discrimination between physical binaries and unrelated optical companions. This is most critical for wide binaries because the probability of interlopers is proportional to the square of the separation. However, many new pairs with separations on the order of 5\arcsec~also need confirmation, particularly near the Galactic plane where the density of stars is high. At this time, we have only been able to collect data on a portion of the companions detected in R15; future observations will cover the other systems. 

Quite unexpectedly, observations with PALM-3000 resolved several subsystems missed by the Robo-AO instrument in R15 because their companions were either too close or too faint. The old rule about new techniques bringing new discoveries is once again verified. 

The observations and data reduction are covered in Section~\ref{sec:obs}. In Section~\ref{sec:ptm} we derive individual magnitudes and colors of the components and place them on the CMD. Each multiple system is commented on in Section~\ref{sec:ind}.  Discussion and conclusions are in Section~\ref{sec:disc}.

%%%%%%%%%%%%%%%%%%%%%%%%%%%%%%%%%%%%%%%%%%%%%%%%

\section{OBSERVATIONS}\label{sec:obs}

We observed the stars on 2013 September 28 UT with the Palomar Observatory Hale 5 m telescope using the PALM-3000 AO system and the PHARO near-IR camera. The PALM-3000 AO system is a natural guide-star system using two deformable mirrors \citep{dekany2013}. One corrects low-amplitude high spatial frequency aberrations, while the other corrects the higher-amplitude low spatial frequency aberrations. It is optimized for high contrast observations and routinely produces Strehl ratios greater than 80\% in \textit{Ks} band. The PHARO camera uses a HgCdTe HAWAII detector for observations between 1 and 2.5 \micron ~wavelength \citep{hayward2001}. The camera has multiple filters in two filter wheels. The filter wheels contain spectral band filters and neutral density filters. These neutral density filters are used for observing bright stars, which saturated the detector. Unfortunately, both the 1\% and the 0.1\% ND filter cause ghost reflections which appear as stellar companions. These ghosts always appear in the same locations and during analysis they are considered to be part of the point spread function. 

Fifty frames were collected of each object with an exposure time of 1.416 s, the minimum exposure for PHARO. After the observing run, the individual frames were reduced by debiasing, flat fielding, bad pixel correction and background subtraction. Then we created five images by co-adding 10 frames into each image. Creating multiple images allowed us to evaluate the precision of the measurements. The \textit{fitstars} algorithm was used to measure the astrometry and photometry of the objects \citep{tenBrummelaar1996,tenBrummelaar2000}.

We observed six calibration binaries\footnote{\url{http://ad.usno.navy.mil/wds/orb6/orb6c.html}} on the same night as the science targets. We compared their measured astrometry with the ephemeris predicted from their orbits and used the results to compute the plate scale and the position angle offset, 24.9$\pm$0.2 mas pixel$^{-1}$ and 0.7$\pm$0.5$^\circ$ error.   The largest error term is the position angle offset.   In speckle and AO work on binaries, the measurement errors in tangential and radial directions are usually the same or similar.  For the position angle error bar, we quadratically added the position angle  offset error with the product of one radian and the ratio of the measurement error to the measured separation.  The separation error bar was computed simarily, with the measurement error summed quadratically with the product of the plate scale error and the separation. Photometric error bars were assigned using the technique described in \citet{roberts2005}. The resulting photometry and astrometry are presented in Table \ref{results}. The table gives the Washington Double Star (WDS; \citealt{WDS}) Catalog designation for the system as are the HD and HIP numbers for the primary star. This is followed by the discoverer designation, the observation epoch, the astrometry and the photometry.  New discoveries are marked with a footnote to the WDS ddesignation.

% Astrometry
\begin{deluxetable*}{lrrlccccc}
\tabletypesize{\scriptsize}
\tablewidth{0pt}
\tablecaption{Astrometry and Photometry\label{results}}
\tablehead{\colhead{WDS} &\colhead{HD}&\colhead{HIP}&\colhead{Discoverer}& \colhead{Epoch}&\colhead{$\theta$ (\degr)} & \colhead{$\rho$ (\arcsec) } & \colhead{$\Delta$J} & \colhead{$\Delta$Ks}}
\startdata
00487+1841& 4655  & 3795 &RAO 5 Ca,Cb & 2013.8371 & 320.9$\pm$1.0 & \phn0.66$\pm$0.01 &... & 0.86$\pm$0.02\\
%00293-0555& 2638  & 2350 &RAO 1 Ba,Bb &  & 167.29$\pm$1.325 & \phn0.52$\pm$0.010 & 2.50$\pm$0.042 & 1.99$\pm$0.05 \\
01027+0908& 6152  & 4878 &RAO 39 AB & 2013.8371 & \phn37.0$\pm$0.5 & \phn2.81$\pm$0.01 & 5.75$\pm$0.20 & 5.20$\pm$0.06 \\
01075+4116& 6611  & 5276 &RAO 40 &  2013.8371 & 336.0$\pm$0.7 & \phn6.16$\pm$0.02 & 5.4\phn$\pm$0.2\phn & 4.75$\pm$0.03 \\
%02370+2439&   & 12189 &RAO 8 Ba,Bb &  & 284.77$\pm$0.614 & \phn0.55$\pm$0.008 & 3.44$\pm$0.029 & 2.84$\pm$0.03 \\
02462+0536& 17250  & 12925 &RAO 9 &  2013.8370 & 253.7$\pm$0.6 & \phn1.89$\pm$0.01 & 3.26$\pm$0.05 & 2.63$\pm$0.02 \\
03390+4232& 22521  & 17022 &RAO 47 &  2013.8370 & 142.8$\pm$0.6 & \phn1.76$\pm$0.01 & 6.26$\pm$0.83 & 5.74$\pm$0.16 \\
03401+3407\tablenotemark{a}& 22692  & 17129 &RBR 26 Aa,Ab& 2013.8370 & \phn44.0$\pm$1.3 & \phn0.50$\pm$0.01 & 4.2\phn$\pm$0.8\phn & 3.83$\pm$0.26 \\
03401+3407& 22692  & 17129 &STF 425 AB & 2013.8370 & \phn60.8$\pm$0.6 & \phn1.93$\pm$0.01 & 0.16$\pm$0.01 & 0.16$\pm$0.01 \\
03413+4554& 22743  & 17217 &RAO 48AC & 2013.8370 &269.3$\pm$0.6 & \phn4.78$\pm$0.03 & ... & 4.78$\pm$0.04\\
04363+5502& 28907  & 21443 &RAO 35 &  2013.8370 & \phn14.8$\pm$0.7 & \phn5.71$\pm$0.05 & 3.88$\pm$0.07 & 3.04$\pm$0.02 \\
05247+6323& 34839  & 25300 &STF 677 A,Ba & 2013.8370 & 117.9$\pm$0.5 & \phn1.10$\pm$0.01 & ... & 0.57$\pm$0.01 \\
05247+6323\tablenotemark{a}& 34839  & 25300 &RBR 27 Ba,Bb & 2013.8370 & \phn68.0$\pm$4.4 & \phn0.13$\pm$0.01 & ... & 1.25$\pm$0.1\phn \\
05247+6323& 34839  & 25300 &RAO 36 AC & 2013.8370 & 227.0$\pm$0.5 & \phn6.99$\pm$0.01 & ... & 4.35$\pm$0.04 \\
06335+4822& 46013  & 31267 &RAO 80 AC & 2013.8371 & 222.8$\pm$0.5 & \phn4.82$\pm$0.01 & 4.20$\pm$0.15 & 3.28$\pm$0.02 \\
06562+4032\tablenotemark{a}& 50720  & 33355 &RBR 28 Aa,Ab &2013.8371 &125.7$\pm$0.6 & \phn1.50$\pm$0.01 & 6.81$\pm$0.8\phn & 6.07$\pm$0.1\phn \\
06562+4032& 50720  & 33355 &RAO 56 AC & 2013.8371 & 158.0$\pm$0.5 & \phn5.58$\pm$0.01 & 6.02$\pm$0.34 & 5.36$\pm$0.06 \\
17422+3804\tablenotemark{a}& 161163 & 86642 &RBR 29 Aa,Ab &2013.8371 & 202.0$\pm$1.0 & \phn0.08$\pm$0.01 & 1.77$\pm$0.10 & 1.36$\pm$0.5\\
17422+3804& 161163 & 86642 &RAO 20 AB &   2013.8371 & 302.2$\pm$0.3 & \phn2.23$\pm$0.01 & 4.07$\pm$0.13 & 3.60$\pm$0.03 \\
19359+5659& 185414 & 96395 &RAO 87 &  2013.8371 &251.3$\pm$0.5 & 10.04$\pm$0.02 & 4.63$\pm$0.15 & 3.74$\pm$0.03 \\
20312+5653& 195872 &101234 &RAO 22 &  2013.8367 &163.8$\pm$3.4 & \phn0.17$\pm$0.01 & 2.48$\pm$0.3\phn & 2.27$\pm$0.18 \\
20333+3323& 195992 &101430 &HJ 1535 A,Ba &2013.8368  & 246.3$\pm$0.5   & 17.05$\pm$0.03 & 3.60$\pm$0.05 & 2.87$\pm$0.02 \\ 
20333+3323\tablenotemark{a}& 195992 &101430 &RBR 29 Ba,Bb & 2013.8368 & 294.5$\pm$9.1 & \phn0.17$\pm$0.03 & 0.09$\pm$0.02 & 0.25$\pm$0.04\\
20333+3323& 195992 &101430 &RAO 71 AE & 2013.8368 &225.8$\pm$0.5 & 12.13$\pm$0.02 & 5.39$\pm$0.2\phn & 4.60$\pm$0.03 \\
20577+2624& 199598 &103455 &RAO 24 &  2013.8368 & 100.1$\pm$1.0 & \phn0.66$\pm$0.01 & 3.34$\pm$0.50 & 2.79$\pm$0.08 \\
21102+2045& 201639 &104514 &RAO 25&  2013.8368 & 210.0$\pm$0.5 & \phn3.28$\pm$0.01 & 3.91$\pm$0.07 & 3.33$\pm$0.02 \\
21585+0347& 208776 &108473 &RAO 73 &  2013.8368 & \phn89.8$\pm$0.5 & 12.35$\pm$0.02 & ... & 4.44$\pm$0.03 \\
23588+3156\tablenotemark{a}& 224531 &118213 &RBR 30 Aa,Ab &2013.8369 & 348.0$\pm$2.2 & \phn0.41$\pm$0.01 & 5.1$\pm$1.0 & 4.40$\pm$0.50\\
23588+3156& 224531 &118213 &RAO 76&  2013.8369 &\phn87.9$\pm$0.5 & \phn4.84$\pm$0.01 & 5.51$\pm$0.25 & 4.83$\pm$0.04\\
23588+3345& 224543 &118225 &RAO 77 &  2013.8372 & 173.6$\pm$0.6 & \phn5.05$\pm$0.02 & 5.84$\pm$0.21 & 5.25$\pm$0.05 \\
\enddata
\tablenotetext{a}{New Discovery}
\end{deluxetable*}

\section{COLOR-MAGNITUDE DIAGRAM}\label{sec:ptm}

Using the differential photometry in the $Ks$ band from Table \ref{results} and the $i'$ band data from R15, we calculated the individual magnitudes of stellar components. The combined $J$ and $Ks$ magnitudes of close binaries were taken from the 2MASS. However, the SDSS $i'$ magnitudes \citep{Fukugita1996} are not available for all our targets. We interpolated them from the $J$ and $V$ magnitudes using the colorindex $c = V-Ks$ as argument. Taking as a template the 1 Gyr isochrone from the Dartmouth stellar models \citep{Dotter2008}, we approximated the color-color relations for dwarfs less massive than 1.5\,${\cal M}_\odot$ by quadratic polynomials
\begin{eqnarray}
V - i' = -0.235 + 0.3103 c + 0.02118 c^2 , \\
i' - J = 0.435 + 0.2354 c + 0.02414 c^2 . 
\label{eq:colors}
\end{eqnarray}

The two estimates of $i'$ for our targets derived from $V$ and $J$ agree very well, with an RMS difference of only 0.026\,mag. We then averaged these estimates to get the combined $i'$ magnitudes.

\begin{figure}[thb]
    \centering
    \includegraphics[height=6cm]{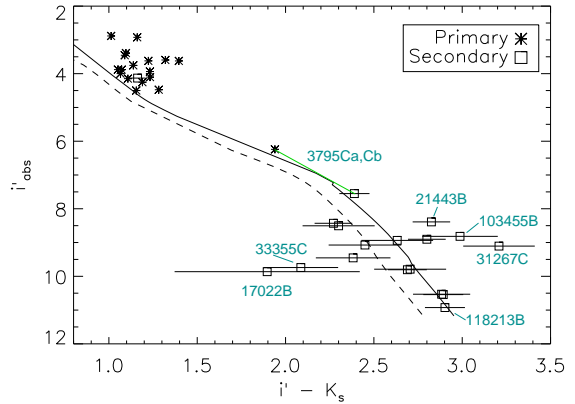} 
    \caption{A color-magnitude diagram for the stars in this paper. The full and dashed lines show the 1 Gyr isochrones for solar metallicity and [Fe/H]=$-$0.5, respectively \citep{Dotter2008}. The green line connects the Ca and Cb components of HIP 3795. Some primary components undoubtably contain unresolved binaries. The Hipparcos names and binary star designations are listed for some of the points. See Section \ref{sec:ind} for discussion of the individual systems.}
    \label{fig:cmd}
\end{figure}

Figure~\ref{fig:cmd} places  primary and  secondary  components of multiple systems resolved both here  and in R15 in the $(i', i'-Ks)$ CMD.     The    {\it   Hipparcos}    parallaxes    of   the    primary targets \citep{HIP2} are used.  The lines are the 1 Gyr isochrones for solar  metallicity  and   [Fe/H]=$-$0.5  from  the  Dartmouth  stellar models      \citep{Dotter2008}     retrieved     from      the     web interface. \footnote{\url{http://stellar.dartmouth.edu/models/webtools.html}} The errors  of the  color  indices of  secondary components  are shown.  They  are  computed   by  assuming  realistic  errors  of  the differential  $i'$-band  photometry  in  R15, typically  0\fm1,  and combining them  with the  errors reported in  Table~1.  Note  that all primary  components (as  well  as  other F-  and  G-type stars  within 67\,pc) are located above the main sequence, indicating potential bias in the Dartmouth models. Some primaries are in fact close binaries.

%%%%%%%%%%%%%%%%%%%%%%%%%%%%%%%
\section{COMMENTS ON INDIVIDUAL SYSTEMS}\label{sec:ind}

Stellar systems with three or more components require individual analysis. We provide below comments on each object, identified by the {\it Hipparcos} number of the primary component. The location of faint components on the CMD is used jointly with the astrometric information and field crowding to evaluate whether the binary is physical or optical (chance alignment).  For some new pairs we compare in Table~\ref{motion} the differential motion of the binary with the proper motion (PM) of the main target \citep{HIP2}. Differential astrometry from Table \ref{results} is combined with the first epoch taken either from R15 (assuming positional errors of half a pixel or 22\,mas when they are not listed) or from 2MASS (accuracy of 0\farcs1 is assumed). Columns (1) and (2) identify the pair by its {\it Hipparcos} number and components. The time difference $\Delta t$ in Column (3) exceeds 10 years when the first epoch is taken from 2MASS. Columns (4) and (5) give the computed relative motion in right ascension and declination, while Columns (6) and (7) list the PM of the main target. Column (8) gives the number of 2MASS stars expected within the binary separation. These estimates do not take into account the brightness of the companion and the magnitude distribution of background sources, and are just a crude indication of the odds that a binary is optical. The last Column (9) summarizes the status of the companion (O for optical, P for physical, P? for likely physical and ? for status unknown). Background stars usually have a small PM, and in such cases the differential PM is almost opposite to PM(A). On the other hand, a detectable relative motion can be orbital or can be caused by motions in inner subsystems and does not necessarily mean that the binary is optical.

\begin{deluxetable*}{rcr ccr rcl}
\tabletypesize{\scriptsize}
\tablewidth{0pt}
\tablecaption{Relative motion\label{motion}}
\tablehead{
\colhead{HIP}&
\colhead{Comp.}&
\colhead{$\Delta t$} &
\colhead{$\Delta \mu^*_\alpha$ } &
\colhead{$\Delta \mu_\delta$} &
\colhead{ $\mu^*_\alpha$} &
\colhead{$\mu_\delta$} &
\colhead{Number} &
\colhead{Status} \\
 &  &  (yr) & 
\multicolumn{2}{c}{ (mas yr$^{-1}$)} & 
\multicolumn{2}{c}{ (mas yr$^{-1}$)} & 
2MASS & (P/O)
}
\startdata
  3795 &Ca,Cb &   1.2 &     \phn\phn$-$4$\pm$20\phn   &    \phn$-$18$\pm$20\phn   &    $-$13 &    $-$49 & 0.000 & P  \\
  4878 &AB    &   1.2 &    \phn$-$32$\pm$25\phn   &   $-$101$\pm$23\phn   &     66 &    $-$23 & 0.004 & P?  \\
  5276 &AB    &   1.2 &     \phs\phn56$\pm$62\phn   &     \phs\phn17$\pm$36\phn   &    118 &    $-$55 & 0.047 & P  \\
 12925 &AD    &   1.2 &      \phs\phn\phn5$\pm$9\phn\phn   &     \phs\phn28$\pm$16\phn    &     73 &    $-$43 & 0.001 & P  \\
 17022 &AB    &   1.1 &     \phs\phn12$\pm$24\phn   &    \phn$-$11$\pm$22\phn   &   $-$194 &   $-$132 & 0.008 & P?  \\
 17217 &AC    &   1.1 &     \phs\phn48$\pm$32\phn   &     \phs\phn21$\pm$48\phn   &    $-$49 &      2 & 0.059 &  P? \\
 21443 &AB    &   1.1 &     \phs\phn29$\pm$67\phn   &    \phn$-$33$\pm$52\phn   &     $-$5 &     20 & 0.155 & P?   \\
 25300 &AC    &  14.8 &    \phs \phn15$\pm$7\phn\phn    &   \phn$-$20$\pm$7\phn\phn    &   $-$127 &    $-$58 & 0.087 & P   \\
 31267 &AC    &   0.2 &    \phs201$\pm$182  &    \phs766$\pm$175  &     44 &     31 & 0.032 & ?  \\
 33355 &AC    &   0.8 &   $-$125$\pm$64\phn   &   $-$149$\pm$38\phn   &      3 &    $-$37 & 0.048 & ?  \\
 86642 &AB    &   1.1 &     \phn$-$11$\pm$23\phn   &     \phs\phn\phn7$\pm$26\phn   &   $-$109 &     94 & 0.007 & P  \\
 96395 &AB    &  14.8 &     \phn\phn$-$6$\pm$7\phn\phn    &    \phs \phn22$\pm$9\phn\phn    &      0 &   $-$200 & 0.197 & P  \\
101234 &AB    &   1.1 &     \phs\phn88$\pm$21\phn   &      \phs\phn\phn4$\pm$22\phn   &   $-$155 &   $-$142 & 0.000 & P?  \\
101430 &AB    &  13.8 &     \phs\phn\phn8$\pm$8\phn\phn    &      \phs\phn\phn6$\pm$11\phn    &    154 &    140 & 2.610 & P  \\
101430 &AE    &  13.8 &   $-$167$\pm$10\phn    &   $-$140$\pm$10\phn    &    154 &    140 & 1.321 & O  \\
103455 &AB    &   1.1 &     \phs\phn50$\pm$9\phn\phn    &     \phs\phn17$\pm$10\phn    &    273 &     95 & 0.002 & P  \\
104514 &AB    &   1.1 &    \phn$-$33$\pm$29\phn   &     \phs\phn43$\pm$24\phn   &    $-$91 &     67 & 0.021 & P  \\
108473 &AB    &  13.8 &     \phs\phn\phn0$\pm$7\phn\phn    &     \phn\phn$-$4$\pm$12\phn    &   $-$248 &   $-$133 & 0.122 & P  \\
118213 &AB    &   1.2 &     \phn\phn$-$4$\pm$20\phn   &    \phn$-$51$\pm$41\phn   &     76 &   $-$134 & 0.014 & P  \\
118225 &AB    &   1.2 &    \phn$-$51$\pm$49\phn   &    \phn$-$13$\pm$26\phn   &    272 &   $-$130 & 0.022 & P  
\enddata
\end{deluxetable*}

For some of the systems, we estimated the orbital period, $P$, of the binary from the measured separations, $\rho$, distances, and mass sum ${\cal M}$ using Kepler's Third Law, $\frac{a^3}{P^2} = \cal M$. The median ratio between the projected separation, $d$ $( = \frac{\rho}{\pi_{\rm HIP}})$, and orbital semi-major axis, $a$, is close to one \citep{tokovinin2014a}, with scatter by a factor of two caused by orbital phase, orbit orientation, and eccentricity. The strict lower limit is $a > \frac{d}{2}$. Statistical period estimates using the assumption $a=d$ are denoted as $P^*$.

\textit{HIP 3795 (HD 4655 = WDS 00487+1841)} is a quadruple system.  The inner AB pair, BU 495, has an orbit with a 143.6$\pm$4.0 yr period \citep{scardia2000}. The CPM component C  at 152\arcsec from AB was found by \citet{LEP}, who estimate the probability of it being physical as 0.99. The component C
was resolved by R15 into a  0\farcs67 pair Ca,Cb. The PM is too small to determine if the Ca and Cb share common proper motion, but the low density of background stars  and the small separation imply that the pair Ca,Cb is physical.

\textit{HIP 4878 (HD 6152 = WDS 01027+0908)} is a double-lined spectroscopic binary with $P=26.2$\,d \citep{Griffin2003}. The tertiary companion B at 2\farcs8 did not move substantially in a year since its  discovery by R15, but the PM of the main component A is small
and the astrometry is not conclusive.
 B is located below the main sequence (MS) on the CMD, although still within the errors of its photometry.  Considering the small  field crowding, we count B as likely physical.

\textit{HIP 5276 (HD 6611 = WDS  01075+4116)} is a triple system with the inner 74.1 d spectroscopic pair    \citep{Gorynya2014} and the tertiary from R15 confirmed as physical by our  photometry  and astrometry. Our observations demonstrate that  the photometry of  B given in 2MASS is biased by the proximity of the bright primary, as in many other similar pairs. Their ``unusual'' colors derived from the 2MASS photometry are thus wrong.

\textit{HIP 12925  (HD 17250 = WDS 02462+0536)} is a quadruple system with a 3-tier hierarchy: a close spectroscopic pair with yet unknown period, the Robo-AO companion D at 1\farcs89,  and the CPM  companion C  at 494\arcsec~\citep{LEP}.  Yet another star, HIP~12862, at  0\fdg9, also shares the common  PM and parallax. The stars are young and belong to the Tucana-Horlogium moving group \citep{Zuckerman2011}. Our  photometry places  D slightly above the  MS.  The companion D was  independently discovered by \citet{Brandt2014} by high-contrast  imaging  at Subaru.   They  measured  it  on 2012.0  at 252\fdg9 and 1\farcs893, in excellent agreement with R15. 

\textit{HIP 17022 (HD 22521 = WDS  03390+4232)}  is a triple system consisting of an astrometric binary with  period  of  3\,yr \citep{Goldin2007}  and  the  Robo-AO tertiary  B  at 1\farcs76.   Considering the  fast PM(A), the fixed position of AB during 1 year would appear to confirm it as a physical binary.   B is located well  below the MS, although its photometry has a large uncertainty. If it is  a physical component with unusual colors, the system merits further study.

\begin{figure*}[htb]
 \centering
 \begin{tabular}{ccc} 
\includegraphics[height=4cm]{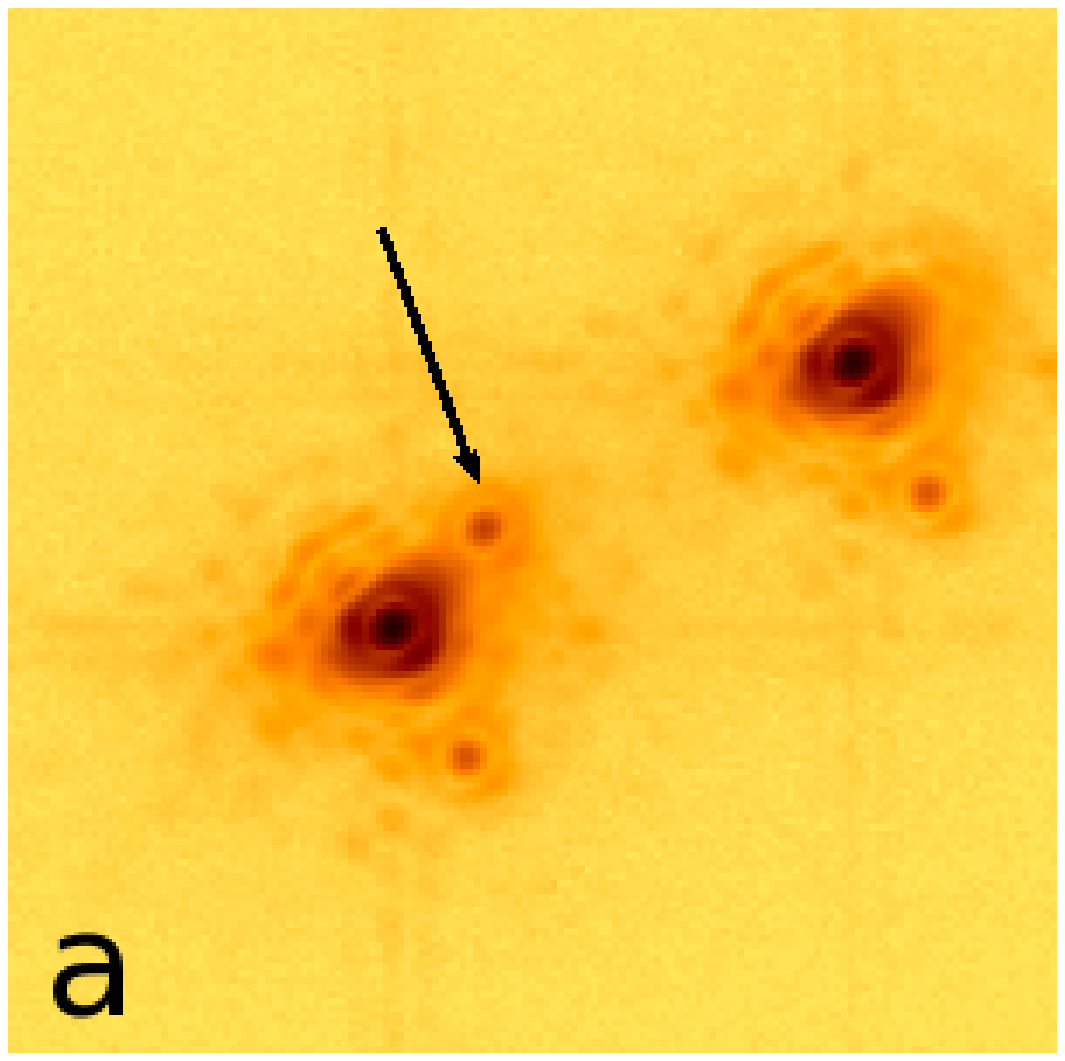} 
\includegraphics[height=4cm]{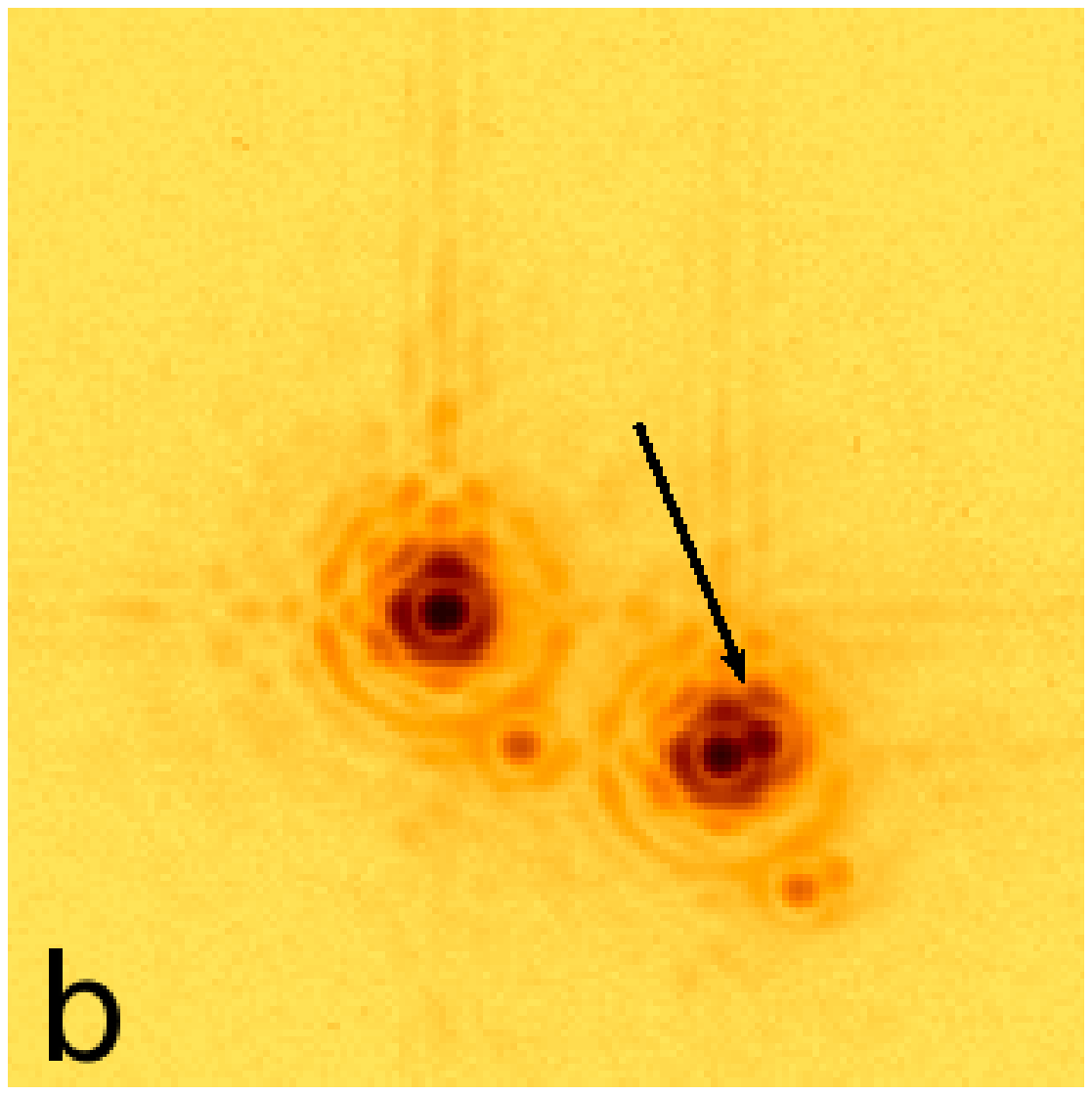} 
\includegraphics[height=4cm]{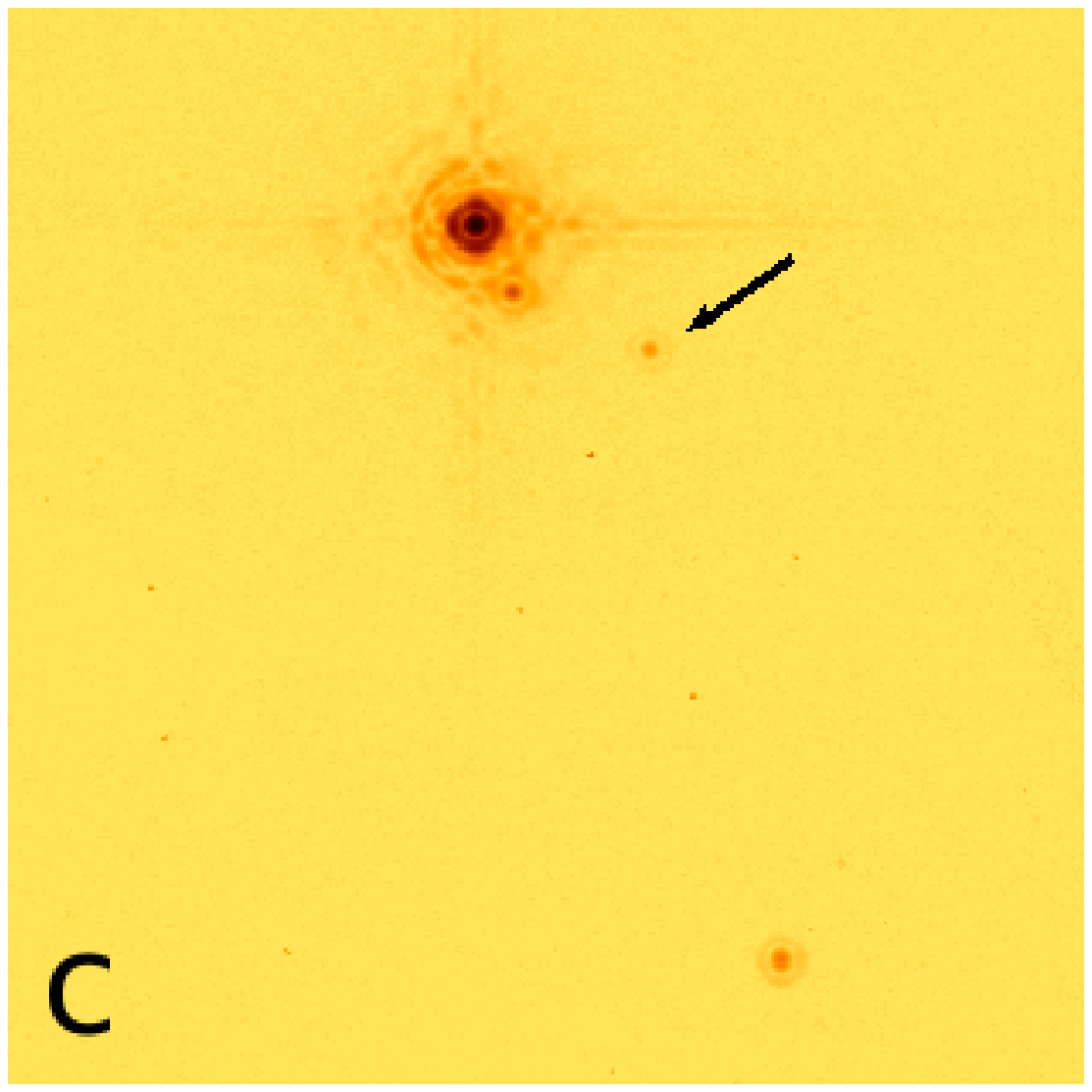} \\
\includegraphics[height=4cm]{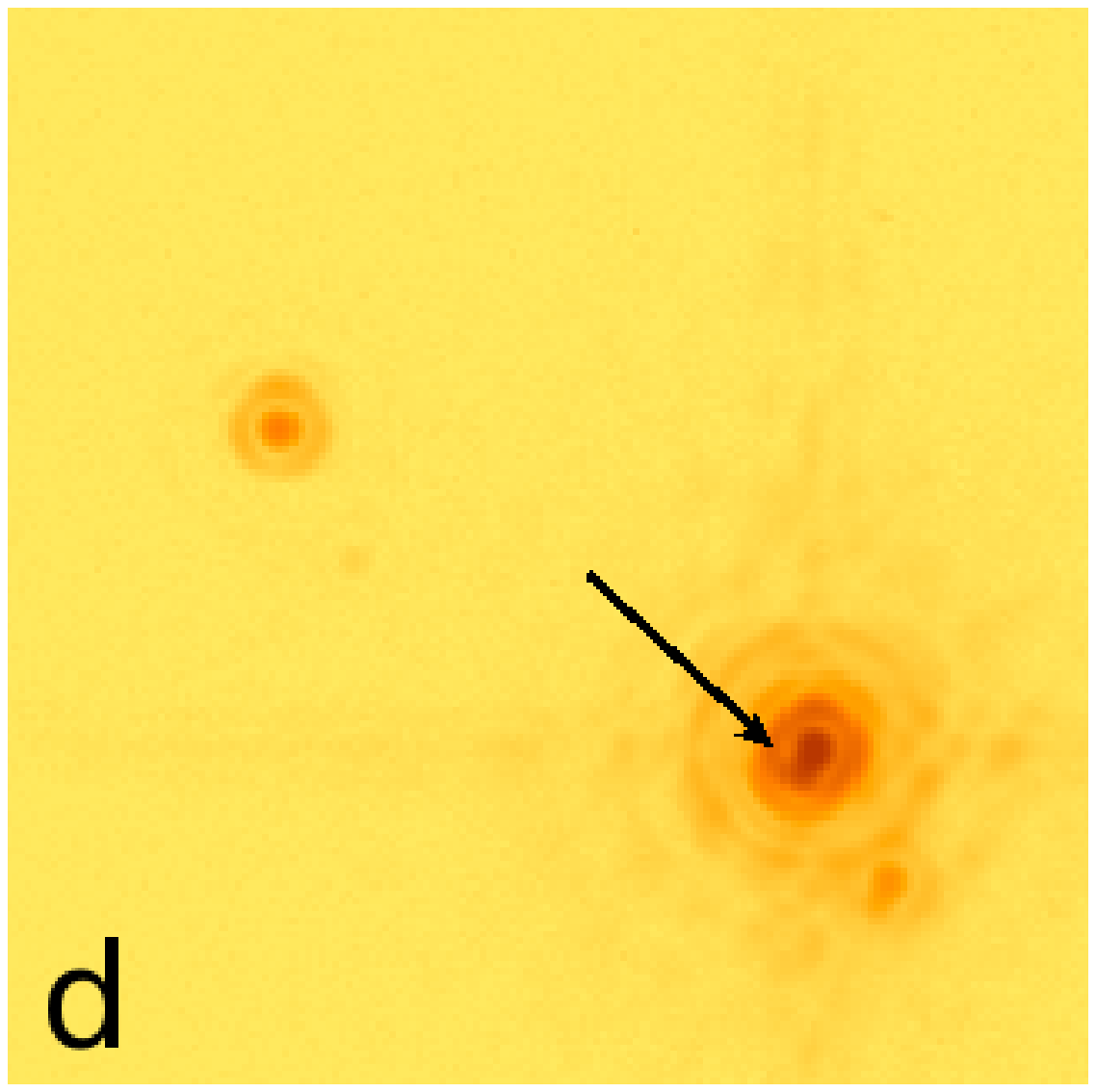} 
\includegraphics[height=4cm]{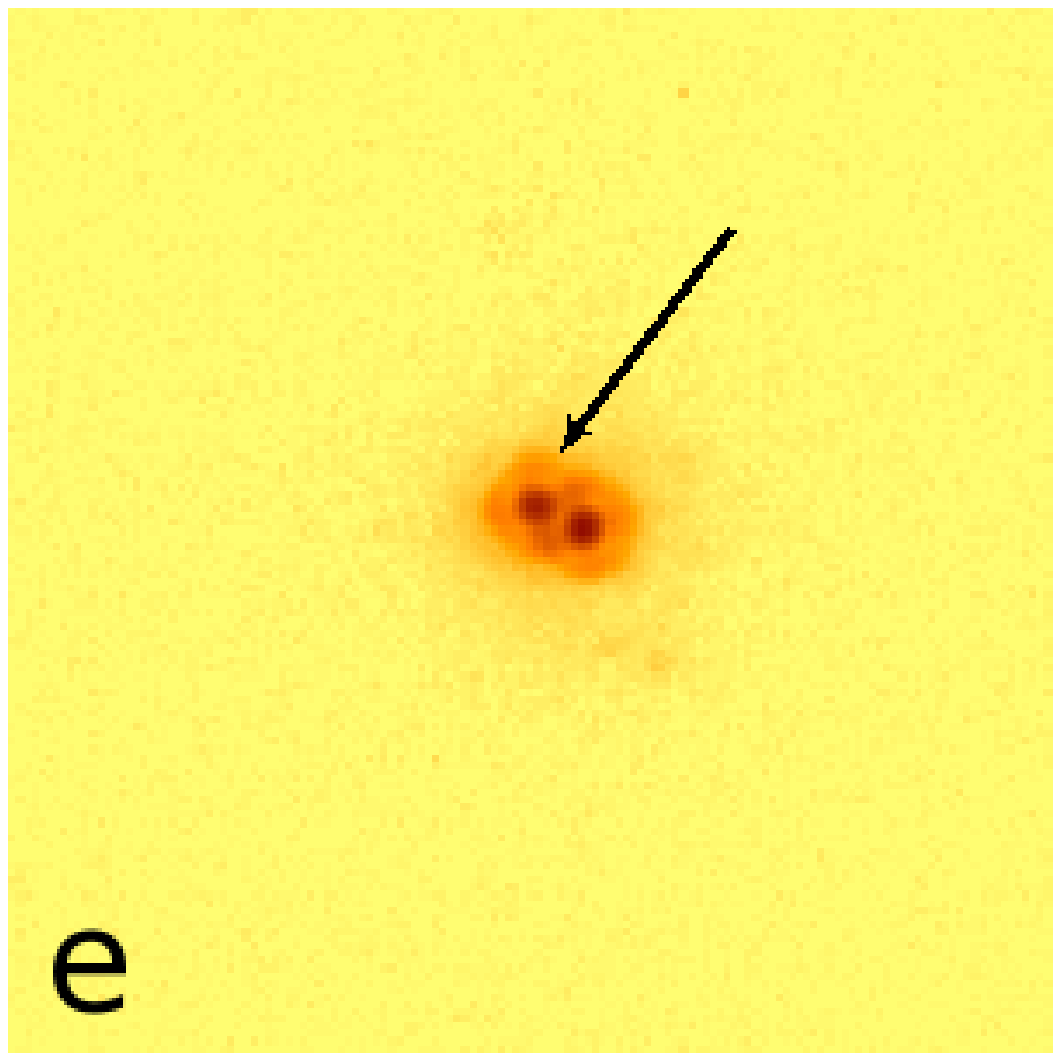} 
\includegraphics[height=4cm]{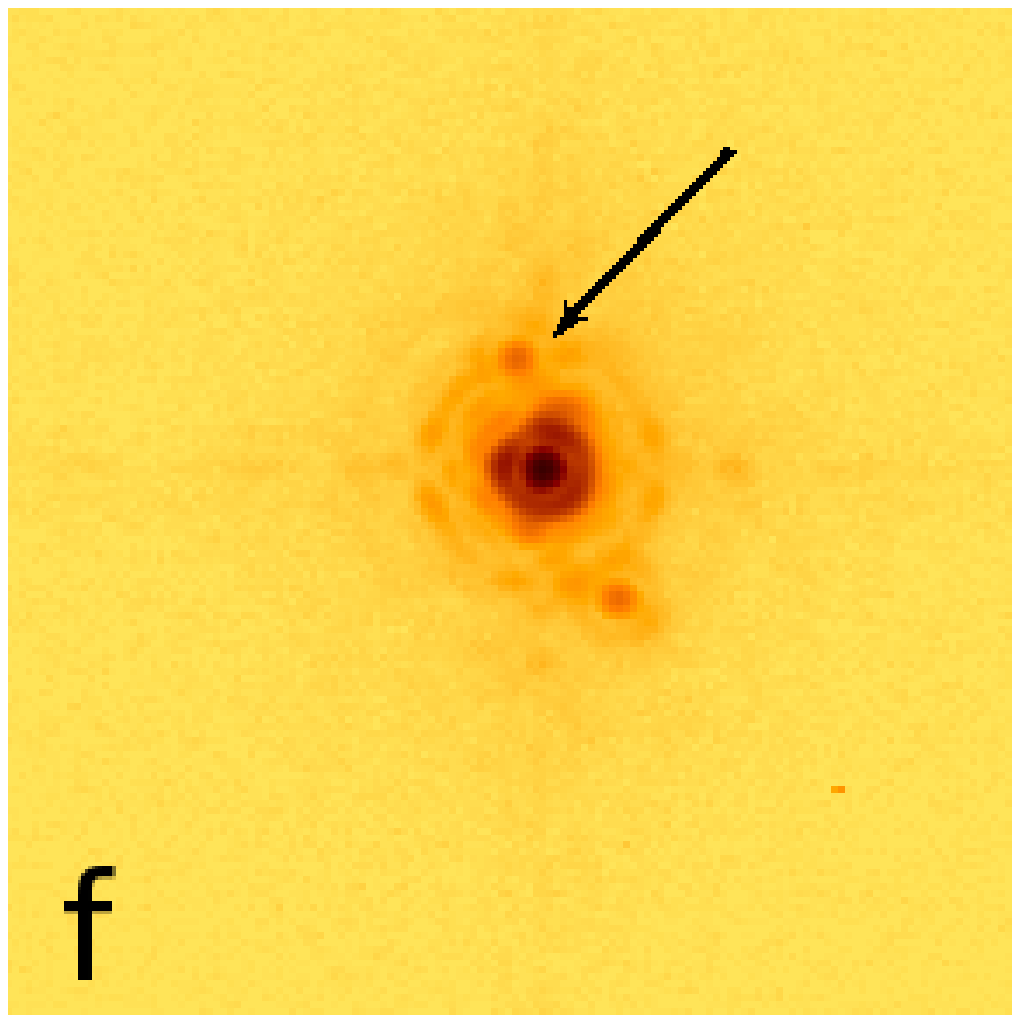} 
 \end{tabular}
 \caption{Images of the six new components detected. North is up and East is to the right. The images are: a) HIP 17129 , b) HIP 25300, c) HIP 33555, d) HIP 86642, e) HIP 101430, f) HIP 118213. Black arrows point to the locations of the newly imaged companions.  The arrows have lengths of one arcsecond. The images were all acquired with the $Ks$ filter on 2013 September 28 UT. In each of the images, there is a ghost to the lower right of each star caused by a neutral density filter in the PHARO camera. These are subimages from the actual data frames; the field of view varies and was chosen to best display the binary. The images are all stretched to best display the binary and the PSF structure. \label{new_binaries} 
}
 \label{images}%
\end{figure*}

\textit{  HIP 17129  (HD 22692 = WDS  03401+3407)}  is  a known  binary,  STF~425, which  shows apparently a non-Keplerian motion. \citet{STF425} proposed that the motion was caused by an unseen companion and calculated an astrometric orbit for it with $P=107$\,yr, eccentricity 0.61, and photocenter semi-major axis 0\farcs179. Our observations were able to resolve the predicted subsystem Aa,Ab.   For the time of our  observations the orbit  predicts a companion position angle of 43\fdg3, in excellent  agreement with  the position  angle  of 44$\pm$1.3$^\circ$ measured here.   The  ratio  of the  predicted  displacement to  the measured separation of  0\farcs50 is $r = 0.32  = q/(1+q)$. Hence, the mass  ratio $q=0.47$ (the  contribution of  the light  from Ab  to the photo-center is  neglected). The  mass of Aa  is estimated  at 1.16\,$ {\cal  M}_\odot$ from its absolute magnitude,  leading  to  0.54\,${\cal M}_\odot$  for  Ab.   The absolute magnitude of Ab corresponds to a smaller mass of 0.33\,${\cal M}_\odot$.   This discrepancy, if  confirmed, can  be explained  by Ab being  a close pair  of M-dwarfs,  as happens  in  other known multiple systems, e.g. in $\kappa$~For \citep{kappaFor}.  See Figure \ref{new_binaries}a for the discovery image. 

\textit{HIP 17217  (HD  22743 = WDS 03413+4554)} is a known triple system. The inner binary has been known for over a century \citep{burnham1894} and its first orbit was computed by R15.  R15 also detected a third component C which was confirmed by examination of 2MASS images.  The close binary is detectable in our $Ks$ images, but the PSFs of the stars are overlapping and we are unable to make a consistent measurement with $fitstars$.  We are able to extract the astrometry of the AC pair and consider it as likely physical.

\textit{  HIP 21443 (HD 28907 = WDS  04363+5502)}  is triple, consisting  of a  2.6-d spectroscopic binary  \citep{Gorynya2014} and  the Robo-AO  and 2MASS  companion at 5\farcs7. The field  is crowded  and  PM(A) is small. Our photometry  places B  slightly above  the MS.  The star  is  young and active according to \citet{Guillout2009} and we retain B as likely physical.

\textit{HIP 25300 (HD 34839 = WDS  05247+6323)}  is now a 3-tier quadruple system since we resolved the secondary component  of the binary AB  = STF 677 into  a new 0\farcs18 pair Ba,Bb and  confirmed that the Robo-AO tertiary  C at 7\arcsec~is physical. We find C to be  slightly below the MS,  within errors. Although the relative motion of AC is fast, it is 
not directed away from PM(A) as would be the case for a distant background star; it  is produced by the orbital motion of the inner binary AB.  The presence of an inner subsystem in the binary AB was suspected previously from its variable RV \citep{N04}.  The  estimated masses of Ba and Bb, 0.98 and 0.65\,${\cal M}_\odot$ respectively, remove the discrepancy between the mass sum of 3.05\,${\cal M}_\odot$ calculated for  the AB system from its grade-4 orbit  by \citet{Hrt2008} and  the expected  mass sum.   Although such discrepancies are not  uncommon for low-quality orbits, in  this case it serves as an  indirect confirmation of Ba,Bb. The  estimated period of Ba,Bb is on  the order of 20\,yrs. See Figure \ref{new_binaries}b for the discovery image. 

\textit{HIP 31267  (HD 46013 = WDS  06335+4822)} is another triple where the inner spectroscopic binary with a period of  1.3\,yr (D.~Latham, 2012, private communication) has a tertiary  companion  discovered with  Robo-AO and seen in the archival 2MASS image. This component is located above the MS, but its photometry   has  large  errors.    Its status remains indeterminate because the PM(A) is only 0\farcs053~yr$^{-1}$ and we do not  have a sufficiently  long time  base. 
A wider pair (AB = UC 1450 at 53") was identified as a common proper motion pair \citep{hartkopf2013}, however, the colors we find here implies that it is optical. The  crowding is moderate.   Possibly the preferred motion of background stars accidentally matches the slow PM of A.   

\textit{  HIP 33555  (HD 50720 = WDS  06562+4032)}  has a  slow  PM of  0\farcs037~yr$^{-1}$.  The R15  companion  at 5\farcs6  is located below  the MS  and   could be  optical, despite moderate crowding.  We discovered  another similarly faint star at 1\farcs9 which  was not spotted in the Robo-AO $i'$-band image. Little can be said about its status. The main target itself is a close spectroscopic  binary (D.~Latham, 2012, private communication) and an X-ray source. See Figure \ref{new_binaries}c for the discovery image. 

\textit{  HIP 86642  (HD  161163 = WDS 17422+3804)} is  a triple  system.   The R15 companion at 2\farcs2  is confirmed  as  physical  by its  fixed  position and  its location on  the CMD.  The main star is a  double-lined spectroscopic binary  with $P=6$\,yr  (D.~Latham, 2012,  private communication) and estimated  semi-major  axis  of  0\farcs1.  Our standard algorithm, $fitstars$, produced subpar results on this stars and instead we used the deconvolution technique used in \citet{riddle2015} to analyze this star.  It is  resolved  here  at 0\farcs07. The magnitude difference  of  Aa,Ab  ($\Delta  K =  2.43$) matches the spectroscopic mass ratio of 0.52 and corresponds to the Ab mass of  0.60\,${\cal M}_\odot$.  The  orbital motion of Aa,Ab can be followed with AO and speckle interferometry. See Figure \ref{new_binaries}d for the discovery image. 

\textit{  HIP  96395  (HD 185414 = WDS 19359+5659)} is  triple.   Its  inner  system has  a  preliminary spectroscopic period  of 14\,yr and a low  amplitude (D.~Latham, 2012, private  communication).   Despite the  estimated  semi-major axis  of 0\farcs25,  the spectroscopic secondary  is too  faint for  its direct resolution without a high-contrast coronagraph. The 2MASS companion at 10\arcsec, noted first by \citet{Fuhrmann2004}, shares the large PM of A and is located on the MS.

\textit{HIP 101234  (HD 195872 = WDS  20312+5653)} is an acceleration binary \citep{MK05} and was first resolved by R15.  The position angle in this paper measured on 2013.8367 differs by 27$^\circ$ from the 2012.7592 measurement.  That is not the only discrepancy between the two data sets.  The photometry with P3K shows large magnitude differences ($\Delta J  = 2.5$ and  $\Delta  K  = 2.3$),  while  Robo-AO  detected  a binary  of  equal brightness. This system is not plotted in Figure~\ref{fig:cmd}.   The separation between the two components is fairly small and in the P3K data, the companion lies on the Airy ring of the primary, which could increase the errors.  The primary is a high proper motion star,  the pair AB   is most likely a physical binary.  Additional observations are necessary to understand this system.  
 
\textit{HIP 101430  (HD  195992 = WDS  20333+3323)}  is a new 2+2 quadruple  system.  The outer 17\arcsec ~binary HJ~1535 AB has been known since  1828  \citep{herschel1831}. If B were a background star, the PM of A, 0\farcs208 yr$^{-1}$, would moved it by 38\arcsec ~in 185\,yr. The pair AB is therefore undoubtedly physical. The discrepant PM of B reported in the WDS  could be caused by the subsystem Ba,Bb discovered here.   The  main component  A  has  an astrometric subsystem Aa,Ab with   $P=3.9$\,yr  \citep{Goldin2007}, confirmed spectroscopically by D.~Latham. The component B was  tentatively resolved  by  Robo-AO,  but was thought to be too uncertain to publish in R15.   Now Ba,Bb is  clearly resolved  at 0\farcs17.  Its estimated period is on the order of 30\,yr. The field is extremely crowded   and another faint star E  is detected at 12\farcs1  from A by both Robo-AO and PALM-3000. Comparison with 2MASS shows relative motion opposite to PM(A), so AE is optical. See Figure \ref{new_binaries}e for the discovery image. 

\textit{HIP 103455  (HD  199598 = WDS  20577+2624)} is a new  0\farcs6 binary detected by Robo-AO. It was targeted   because   of   its   variable  velocity \citep{N04}  and   astrometric acceleration \citep{MK05}.   The binary  is confirmed  here, helped  by  the large PM of the primary.  Our  photometry places B  above the main  sequence.  According to \citet{Guillout2009},  the target is young, so  its companion could indeed be a PMS star.  The object was already targeted by \citet{metchev2009} in their survey of young stars,  but the binary was not detected.  After learning of our discovery, Metchev was able to extract a measurement of the star from the archival data (Private Communication S. Metchev).

\textit{  HIP  104514  (HD  201639 = WDS  21102+2045)}  has  a  variable RV \citep{N04},  so  the  3\farcs3  companion B discovered by Robo-AO  makes it a triple system  (the estimated period of AB is 2000\,yr, too long to cause the RV variation). We found the component B  to be  on the [Fe/H]=$-0.5$ isochrone, in agreement with  [Fe/H]=$-0.54$ measured for the star A \citep{Reddy2003}.

\textit{  HIP   108473   (HD 208776 = WDS  21585+0347)} is  a  single-lined   spectroscopic  binary  with $P=7.18$\,yr    \citep{Nidever2002}     and    a    large     PM    of 0\farcs28~yr$^{-1}$.   The   distant  component  B   at  12\farcs3  is confirmed as physical by 2MASS,  Robo-AO, and this work. It is located slightly  below the  MS, just  like  HIP~104514B. The  object is  also metal-deficient   relative    to   the   Sun,    [Fe/H]=$-0.14$   \citep{Zielke1970}.

\textit{HIP 110574 (HD 212426 = WDS 22240+0612)} is a triple system consisting of a close binary Aa,Ab with a separation of 0\farcs09 detected by R15 and a CPM companion with a separation of 171\arcsec.  The Aa,Ab system is unresolved in the $Ks$ images, but is elongated in the $J$ images. We were unable to extract astrometry from the data. The separation has decreased since it was discovered in 2012; \citep{tokovinin2014c} was able to resolve it with  visible speckle interferometry at the 4.1 m SOAR telescope on 2013.73 at 41\,mas \citep{tokovinin2014c}.  

\textit{HIP 118213  (HD 224531 = WDS 23588+3156)} was previously thought to be a binary, but our observations reveal that it is actually a triple system.  The outer component B at 4\farcs8  was discovered by Robo-AO and confirmed by the 2MASS  image and by the new data presented  here. It is located near  the low end of the MS  in  Figure~\ref{fig:cmd}.  The main  star  with astrometric acceleration \citep{MK05} is now  resolved at 0\farcs4. This separation corresponds  to an orbital  period on  the order  of 100\,yrs.  The new component Ab  is too  faint for  its detection in  the $i'$  band with Robo-AO. See Figure \ref{new_binaries}f for the discovery image. 

\textit{  HIP  118225  (HD 224543 = WDS  23588+3345)} is a triple system.   The  inner  spectroscopic binary  has $P=25.4$\,d \citep{Latham2002}, the  Robo-AO companion at 5\arcsec~ is seen at  a constant position despite  PM(A)= 0\farcs3~yr$^{-1}$, while its color places it on the MS.

%%%%%%%%%%%%%%%%%%%%%%%%%%%%%%%%%%%%%%%%%%%%%%%%%%%%%%%%%%%%%%%

\section{SUMMARY}\label{sec:disc}

This study shows the power of high-resolution AO imaging for the study of hierarchical multiplicity. We resolved for the first time inner subsystems in six binaries. Four of those (HIP 17129, 86642, 118213, 25300) had previous indications of subsystems from variable RV and/or astrometric acceleration. The AO discovery space overlaps with these alternative techniques, but direct resolution of subsystems allows their characterization in terms of period estimated from separation and mass ratio estimated from photometry. As the orbital periods of these subsystems range from several years to several decades, such characterization by RV or astrometry would necessarily involve a long-term monitoring, which is not available in most cases. The exception is HIP~17129, where a subsystem with $P=107$\,yr, first resolved here, was inferred from the strange motion of the visual pair observed for almost two centuries.

The new subsystem in the secondary component of HIP~101430 was totally unexpected. This 17\arcsec visual pair known for 185 years is now revealed as a 2+2 quadruple system, where each of the visual components is in turn a close binary. The fainter secondary components of visual binaries generally have much less information on subsystems compared to the brighter primaries \citep{tokovinin2014a}, and here high-resolution AO imaging makes a large difference. 

Our strategy of observing binaries to discover subsystems has been proven successful. The emerging statistics \citep{law2010,tokovinin2014b} indicates that a third to a half of visual binaries harbor subsystem(s). The simple fact that binaries can contain more than just two stars is often forgotten or ignored, for example when comparing PMs of components in wide binaries or calculating their orbits from short arcs while assuming single-star masses. Similarly, RV monitoring of wide binaries in search of exo-planets is incomplete without full characterization of subsystems \citep{roberts2015}. 

This work contributes to the statistical characterization of hierarchical multiplicity in the solar neighborhood. It will help to understand the formation mechanisms that create the panoply of single, binary, and multiple stars, and thus to understand the origin of stars and planets in general. 

There is additional work that can be done with these systems.  Additional astrometric measurements will show orbital motion confirming that the systems are physically bound.  In addition, future astrometric measures can be combined with the RV measurements that have been made on several of the systems and will allow for the computation of the full orbital solution.  Though there are strong indications that the newly observed companions are physical, with the exception of HIP 33555, the new companions need to be confirmed with additional observations.   Additional observations of HIP 101234 are need to resolve the discrepancy between the visible and the near-IR differential magnitude.  With a period of only six years measured by RV, HIP 86642 Aa,Ab is a prime target for frequent follow up astrometric measurements that can be combined with the double lined spectroscopic orbit.

%%%%%%%%%%%%%%%%%%%%%%%%%%%%%%%%%%%%%%%%%%%%%%%%%%%%%%%%%%%%%%%

\acknowledgements

We thank D. Latham for the insight he provided on a number of these systems. This paper is based on observations obtained at the Hale Telescope, Palomar Observatory. A portion of the research in this paper was carried out at the Jet Propulsion Laboratory, California Institute of Technology, under a contract with the National Aeronautics and Space Administration (NASA). This research made use of the Washington Double Star Catalog maintained at the U.S. Naval Observatory, the SIMBAD database, operated by the CDS in Strasbourg, France and NASA's Astrophysics Data System. This publication made use of data products from the Two Micron All Sky Survey, which is a joint project of the Universit of Massachusetts and the Infrared Processing and Analysis Center/California Institute of Technology, funded by NASA and the National Science Foundation.

\textbf{ Facilities:} \facility{ \facility{Hale (PHARO)} } 
 
%%%%%%%%%%%%%%%%%%%%%%%%%%%%%%%%%%%%%%%%%%%%%%%%%%%%%%%%%%%%%%%

% References


\begin{thebibliography}{}

\bibitem[Brandt et al.(2014)]{Brandt2014} 
         Brandt, T. D., Kuzuhara, M., McElwain, M.W. et al. 2014, \apj, 786, 1 
 
\bibitem[Burnham(1894)]{burnham1894}
         Burnham, S.W. 1894, Pub. Lick Obs., 2, 206

\bibitem[Dekany et al.(2013)]{dekany2013}
         Dekany, R., Roberts, J., Burruss, R. et al. 2013, \apj, 776, 130

\bibitem[Dotter et al.(2008)]{Dotter2008} 
         Dotter, A., Chaboyer, B., Jevremovi\'c, D. et al. 2008, \apjs, 178, 89

\bibitem[Duch\^ene \& Kraus(2013)]{DK13}
         Duch\^ene, G. \& Kraus, A. 2013, \araa, 51, 269

\bibitem[Duquennoy \& Mayor(1991)]{DM91}
         Duquennoy, A. \& Mayor, M. 1991, \aap, 248, 485 

\bibitem[Fuhrmann(2004)]{Fuhrmann2004} 
         Fuhrmann, K. 2004, AN, 325, 3

\bibitem[Fukugita et al.(1996)]{Fukugita1996}
         Fukugita, M., Ichikawa, T., Gunn, J.E. et al. 1996, \aj, 111, 1748

\bibitem[Goldin \& Makarov(2007)]{Goldin2007} 
         Goldin, A. \& Makarov, V.V. 2007, \apjs, 173, 137

\bibitem[Gorynya \& Tokovinin(2014)]{Gorynya2014} 
         Gorynya, N.A. \& Tokovinin, A. 2014, \mnras, 441, 2316 

\bibitem[Griffin \& Suchkov(2003)]{Griffin2003} 
         Griffin, R.F. \& Suchkov, A.A. 2003, \apjs, 147, 103

\bibitem[Guillout et al.(2009)]{Guillout2009} 
         Guillout, P., Klutsch, A., Frasca, A. et al. 2009, \aap, 504, 829

\bibitem[Hartkopf et al.(2008)]{Hrt2008} 
         Hartkopf, W.I., Mason, B.D. \& Rafferty, T.J. 2008, \aj, 135, 1334

\bibitem[Hartkopf et al.(2013)]{hartkopf2013}
         Hartkopf, W.I., Mason, B.D., Finch, C.T., Zacharias, N., Wycoff, G.L. \& Hsu, D. 2013, \aj, 146, 76

\bibitem[Hayward et al.(2001)]{hayward2001}
         Hayward, T.L., Brandl, B., Pirger, B., et al. 2001, \pasp, 113, 105

\bibitem[Herschel(1831)]{herschel1831}
         Herschel, J.F.W. 1831, Mm. RAS, 4, 331

\bibitem[Latham et al.(2002)]{Latham2002} 
         Latham, D.W., Stefanik, R. P., Torres, G. et al. 2002, \aj, 124, 1144
 
\bibitem[Law et al.(2010)]{law2010}
         Law, N.M., Dhital, S., Kraus, A., Stassun, K.G., \& West, A.A. 2010, \apj, 720, 1727

\bibitem[Makarov \& Kaplan (2005)]{MK05}
         Makarov, V.V. \& Kaplan, G.H., 2005, \aj, 129, 2420

\bibitem[Mason et al.(2001)]{WDS}
         Mason, B.D., Wycoff, G.L., Hartkopf, W.I., Douglass, G.G. \& Worley, C.E. 2001, \aj, 122, 3466 

\bibitem[Metchev \& Hillenbrand(2009)]{metchev2009}
         Metchev, S.A. \& Hillenbrand, L.A. 2009, \apjs, 181, 62

\bibitem[Nidever et al.(2002)]{Nidever2002} 
         Nidever, D.L., Marcy, G.W., Butler, R.P. et al. 2002, \apjs, 141, 503

\bibitem[Nordstr\"om et al. (2004)]{N04}
         Nordstr\"om, B., Mayor, M., Andersen, J. et al. 2004, \aap, 418, 989 

\bibitem[Raghavan et al.(2010)]{R10}
         Raghavan, D., McAlister, H.A., Henry, T.J. et al. 2010, \apjs, 190, 1

\bibitem[Reddy et al.(2003)]{Reddy2003} 
         Reddy, B.E., Tomkin, J., Lambert, D.L., \& Allende Prieto, C. 2003, \mnras, 340, 304

\bibitem[Rica Romero \& Zirm(2014)]{STF425} 
         Rica Romero, F. \& Zirm, H. 2014, IAU Circ. \# 183, 1

\bibitem[Riddle et al.(2015)]{riddle2015}
         Riddle, R. L., Tokovinin, A., Mason, B. D. et al. 2015, \apj, 799, 4

\bibitem[Roberts et al.(2005)]{roberts2005}
         Roberts Jr., L. C., Turner, N.H., Bradford, L.W., et al. 2005, \aj, 130, 2262

\bibitem[Roberts et al.(2015)]{roberts2015}
         Roberts JR., L.C., Tokovinin, A., Mason, B.D., Riddle, R.L., Hartkopf, W.I., Law, N.M. \& Baranec, C. 2015, \aj, In Press

\bibitem[ten Brummelaar et al.(1996)]{tenBrummelaar1996}
         ten Brummelaar T.A., Mason B.D., Bagnuolo, Jr. W.G., Hartkopf W.I., McAlister H.A., Turner N.H. 1996, \aj, 112, 1180

\bibitem[ten Brummelaar et al.(2000)]{tenBrummelaar2000}
         ten Brummelaar T.A., Mason McAlister H.A., Roberts Jr. L.C., Turner N.H., Hartkopf W.I., Bagnuolo Jr., W.G. 2000, \aj, 119, 2403

\bibitem[Scardia et al.(2000)]{scardia2000}
         Scardia, M., Prieru, J.-L. Aristidi, E., \& Koechlin, L. 2000, AN 321, 255

\bibitem[Tokovinin \& L\'epine(2012)]{LEP} 
         Tokovinin, A. \& L\'epine, S. 2012, \aj, 144, 102

\bibitem[Tokovinin(2013)]{kappaFor}
         Tokovinin, A. 2013, \aj, 145, 76

\bibitem[Tokovinin(2014a)]{tokovinin2014a}
         Tokovinin, A. 2014a, \aj, 147, 86 
 
\bibitem[Tokovinin(2014b)]{tokovinin2014b}
         Tokovinin, A. 2014b, \aj, 147, 87 

\bibitem[Tokovinin et al.(2014)]{tokovinin2014c}
         Tokovinin, A., Mason, B.D., \& Hartkopf, W.I. 2014, \aj, 147, 123 

\bibitem[van Leeuwen(2007)]{HIP2}
         van Leeuwen, F. 2007, \aap, 474, 653

\bibitem[Zielke(1970)]{Zielke1970} 
         Zielke, G. 1970, \aap, 6, 206

\bibitem[Zuckerman et al.(2011)]{Zuckerman2011} 
         Zuckerman, B., Rhee, J.H., Song, I. \& Bessell, M.S. 2011, \apj, 732, 61

\end{thebibliography}
\end{document}